\documentclass[a4paper,10pt]{article}
\usepackage[utf8]{inputenc}
\usepackage{graphicx}
\usepackage{color}
\begin{document}

\def\pa{\partial}
\def\vb{{\bf v}}
\def\Fb{{\bf F}}
\def\Kb{{\bf K}}
\def\Bb{{\bf B}}
\def\ol{\overline}
\def\eb{{\bf e}}
\def\omb{{\bf \omega}}
\def\MC{meridional circulation}
\def\DF{differential rotation}
\def\CZ{convection zone}

\title{A theoretical estimate of the pole-equator temperature difference and a possible origin of the near-surface shear layer}         
\author{Arnab Rai Choudhuri \\ Department of Physics \\ Indian Institute of Science 
\\ Bangalore - 560012, India}        
\date{}          
\maketitle 

\begin{abstract}

Convective motions in the deep layers of the solar convection zone are affected by rotation, making the convective heat
transport latitude-dependent, but this is not the case in the top layers near the surface. We use the thermal wind balance
condition in the deeper layers to estimate the pole-equator temperature difference. Surface observations of this temperature
difference can be used for estimating the depth of the near-surface layer within which convection is not affected by
rotation.  If we require that the thermal wind balance holds in this layer also, then we have to conclude that this must
be a layer of strong differential rotation and its characteristics which we derive are in broad agreement with the
observational data of the near-surface shear layer. 
\end{abstract}

\section{Introduction}

Whether the surface temperature of the Sun has any variations with latitude is an interesting question both from theoretical
and observational considerations.  We point out that a simple order-of-magnitude estimate of the pole-equator temperature
variation can be made from the thermal wind balance equation, which is the key equation in the theory of the \MC. Although
this equation is well known in the literature (see, for example, Kitchatinov, 2013; Karak et al., 2015, \S5.2; Choudhuri, 2020) 
and this order-of-magnitude estimate follows easily from this equation,
we have not come across a discussion of this anywhere in the literature of the subject. 
Presumably, it is not generally realized that such a simple order-of-magnitude estimate of the pole-equator temperature difference can be
made and that it has important implications.  This order-of-magnitude analysis gives us a clue 
to understanding an enigmatic finding of helioseismology.  Although
the isorotation contours of solar differential rotation are nearly radial within the body of the convection zone, they bend
towards the equator near the surface, giving rise to a near-surface shear layer (Howe, 2009).  The origin of this layer is
still not fully understood, although there have been efforts to explain this on the basis of numerical simulation (Guerrero
et al., 2013; Hotta, Rempel, and Yokoyama, 2015).  We propose a simple alternative explanation of this layer.

Heat is transported by convection in the outer layers of the Sun from about $0.7 R_{\odot}$ to the solar surface. Convective
motions are expected to be affected by the solar rotation if the convective turnover time $\tau_c$ is more than or comparable to the
solar rotation period, i.e. if
$$\tau_c \geq 25 \; {\rm days}. \eqno(1)$$
Immediately below the solar surface, convection takes place in the form of granules with turnover times of the order of a few
minutes.  Since granular convection does not satisfy the condition (1), we believe that convection below the surface till a depth,
say $D$, will be completely unaffected by rotation.  However, below this depth $D$, convection is likely to take place in the
form of giant cells with much longer turnover times. Numerical simulations indeed show the existence of banana-shaped giant cells
roughly aligned parallel to the rotation axis---see Figure~1 in Brown et al. (2010)
or Figure~3 in Gastine et al. (2014). Heat transport in such
convection cells is clearly affected by rotation.  Since the Coriolis force (which is crucial in producing the banana cells by
acting on moving fluid parcels) is more effective in lower latitudes, we expect the effect of rotation to be more at the lower
latitudes.  Heat transport should be more efficient at higher latitudes, giving rise to a higher temperature at the poles. A hotter
pole will tend to drive a \MC\ equatorward at the solar surface (opposite to what is observed).  This type of circulation driven
by temperature differences on isochoric surfaces (i.e.\ surfaces of constant density) is often referred to as a thermal wind.
It was realized nearly half a century ago that the effect of rotation on convection may make the heat transport in the Sun
latitude-dependent and may produce a thermal wind (Durney and Roxburgh 1971; Belvedere and Paterno 1976).   

The thermal wind is opposed by the centrifugal force term arising out of the differential rotation.  With differential rotation
mapped by helioseismology, one can now calculate this centrifugal force term fairly accurately.  Within the main body of the convection
zone, the centrifugal force term is expected to roughly balance the thermal wind term.  In fact, the centrifugal force term must slightly overpower the
thermal wind term to drive the \MC\ in the correct direction.  This balance is, however, thought to be upset in the layers just below the solar
surface.  It has been argued that the dissipative term, which is negligible inside the body of the convection zone compared to the centrifugal force term, becomes
important in the layer near the surface and balances the centrifugal term, the thermal wind term being negligible there (Hotta, Rempel, and Yokoyama, 
2014; Karak et al.\ 2015). We present a different point of view.  We argue that the thermal wind term becomes even more
important in the layer just below the surface and plays a crucial role in creating the near-surface shear layer. 

On the basis of a mean field model of the differential rotation and the \MC, Kitchatinov and R\"udiger (1994) concluded that
the pole of the Sun has to be about 4 K hotter than the equator.
An input from observations will be very crucial in deciding between the various alternative viewpoints.  R\"udiger (1989, p.\ 79) has
provided a summary of the early efforts in determining the pole-equator temperature difference.  Nearly all the authors reported
an upper limit rather than  an actual measurement.  Such efforts have continued (Kuhn, Libbrect, and Dicke, 1988; Rast, Ortiz, and 
Meisner, 2008). In one of the last investigations we are aware of, Rast, Ortiz, and Meisner (2008) reported an excess temperature of
2.5 K in the polar region at the photospheric level.  If this is true, then it provides a strong support to our theoretical
conjecture, as we shall point out.  We hope that this important issue will be settled observationally in the near future.

An estimate of the pole-equator temperature difference from the thermal wind balance
condition is presented in the next Section.  Then \S3 is devoted to determining
the various characteristics of the near-surface shear layer by assuming that this
balance condition holds in this layer also.  Our conclusions are summarized in \S4.

\section{An order of magnitude estimate of the pole-equator temperature difference}       

The well-known equation driving the \MC\ 
(which is the equation of the azimuthal component of vorticity) can be derived from
basic principles of fluid mechanics.  See, for example, Choudhuri (2020)
for a derivation of this equation.
The two important source terms in this
equation are the centrifugal term and the thermal wind term. The dissipation term
within the convection zone turns out to be negligible compared to the centrifugal term
estimated from helioseismology.  This means that the centrifugal term cannot be
balanced by the dissipation term in the interior of the convection zone and must be
balanced by the thermal wind term in order to maintain a steady \MC.  This leads to the equation
$$r \sin \theta \frac{\pa}{\pa z} \Omega^2 =
\frac{1}{r} \frac{g}{\gamma \, C_V}\frac{\pa S}{\pa \theta}, \eqno(2)$$
where $S$ is the specific entropy (i.e.\ the entropy per unit mass), $z$ is
measured parallel to the rotation axis starting from the equatorial plane
and all the other symbols have their usual meanings.
We may mention that Balbus et al. (2009) constructed a theoretical model of differential
rotation by solving this equation to find $\Omega (r, \theta)$. We are somewhat skeptical
about this approach due to the non-uniqueness of this equation.  Suppose we find a solution
$\Omega^2$ of this equation.   If we add any arbitrary function $f (r \sin \theta)$ to this
solution, it is easy to check that this will still be a solution of (2).  However, we can use (2)
to make an order of magnitude estimate of the pole-equator temperature difference
which we need for balancing the centrifugal term that would arise from the
differential rotation measured by helioseismology.  

We now try to estimate the magnitude of the centrifugal term, as given
by the left hand side of (2), below the shallow layer near the solar surface.
We look at Figure~1 of Howe (2009) giving a map of the solar differential
rotation and consider a vertical straight line starting from the bottom
of the convection zone in the equatorial plane and going upwards.  We note
that $\Omega$ has the value $\Omega_{\rm eq}/ 2 \pi = 460$ nHz at the 
beginning of this line at the bottom of the convection zone and the value
$\Omega_{\rm mid}/ 2 \pi = 420$ nHz where it reaches the bottom of the
near-surface shear layer. It is easy to argue that 
the left hand side of (2) in the interior of
the convection would approximately be equal to
$$r \sin \theta \frac{\pa}{\pa z} \Omega^2 \approx - [\Omega_{\rm eq}^2 - \Omega_{\rm mid}^2]. \eqno(3)$$
Substituting the values of  $\Omega_{\rm eq}$ and $\Omega_{\rm mid}$, we get
$$r \sin \theta \frac{\pa}{\pa z} \Omega^2 \approx - [(460)^2 - (420)^2] \times (2 \pi 10^{-9})^2 \; {\rm s}^{-2}
\approx 1.4 \times 10^{-12}  \; {\rm s}^{-2}. \eqno(4)$$

Next, we make an estimate of the right hand side of (2). We note that the specific entropy of an ideal gas
is given by 
$$S = C_V \ln T  - (\gamma -1) C_V \ln \rho + K, $$
where $K$ is a constant. The entropy difference between the equator and the pole on any isochoric surface (i.e.
a surface of constant $\rho$) is
$$\Delta S = C_V \ln\left( \frac{T_{\rm eq}}{T_{\rm pole}} \right).$$
Taking $\Delta T$ to be the temperature excess of the pole with respect to equator, we have
$$\Delta S \approx - C_V  \frac{\Delta T}{T_S}, \eqno(5)$$
where $T_S$ is the temperature of our isochoric surface and we have made use of the approximation $\ln (1 +x)
\approx x$ for $|x| \ll 1$. Since this entropy difference takes place over an angular separation $\pi/2$,
we have
$$\frac{\pa S}{\pa \theta} \approx - 2 C_V  \frac{\Delta T}{\pi T_S}, \eqno(6)$$
Substituting this in the right hand side of (2), we get
$$\frac{1}{r} \frac{g}{\gamma \, C_V}\frac{\pa S}{\pa \theta}
\approx - \frac{2}{ \pi \gamma} \frac{G M_{\odot}}{(0.85 R_{\odot})^3} \frac{\Delta T}{ T_S}, \eqno(7)$$
where we have taken $r$ to be given by $0.85 R_{\odot}$ corresponding to the middle of the convection zone
and have also used this to calculate $g$.  If we now use the standard values
of solar mass and radius, then we get (taking $\gamma = 1.4$)
$$\frac{1}{r} \frac{g}{\gamma \, C_V} \frac{\pa S}{\pa \theta} \approx 2.8 \times 10^{-7} 
\frac{\Delta T}{T_S} \; {\rm s}^{-2}.
\eqno(8)$$

We note that (4) gives the value of the left hand side of (2) only inside
the convection zone underneath the near-surface layer, whereas (8) gives
the value of the right hand side of (2) for any isochoric surface, provided $\Delta T/T$ is much smaller than 1. By equating (4) and (8), we arrive at
$$\frac{\Delta T}{T_S} \approx 5.0 \times 10^{-6}. \eqno(9)$$
If we take $T_S$ equal to the temperature 5800 K at the photospheric surface,
then we get a rather low value $\Delta T \approx 2.9 \times 10^{-2}$  K. But,
should we use the photospheric temperature for $T_S$ in (9)? As we point
out, (4) gives the magnitude of the centrifugal term underneath the
near-surface shear layer.  For the sake of consistency, we may expect
(8) would be equal to (4) only if use $\Delta T/T$ for an isochoric
surface below the near-surface shear layer.  At what depth this isochoric
surface should be is discussed in the next Section.
We stress the rather non-intuitive fact that a centrifugal force
resulting from a significant variation in $\Omega$ needs a very small
pole-equator temperature difference to give rise to a thermal wind term to
balance it.

\section{Physics of the near-surface layer}

As we move from a region inside the convection zone towards the solar
surface, the convection cells would be of smaller size due to decrease in
the pressure scale height and rotation would have less effect on the
convection cells.  While this is a gradual transition, we can make the
following simplification. Below a certain depth $D$, we assume that (1)
is satisfied and convection is significantly affected by rotation.  On
the other hand, (1) is not satisfied above $D$ and convection is unaffected
by rotation.  We now make an estimate of this depth $D$.

Within the convection zone, the temperature gradient $dT/dr$ is very nearly
equal to the adiabatic gradient, the small difference between the two
depending on the mixing length $l$ (Kippenhahn and Weigert, 1990, \S7). 
Since the mixing length $l$ would be latitude-dependent below $D$ due to
the effect of rotation, the gradient $dT/dr$ would vary with latitude
beneath $D$.  On the other hand, the mixing length $l$ would be independent
of latitude above $D$, leading us to conclude the $dT/dr$ would be the
same at all latitudes in this layer near the surface.  This means that the
temperature would fall radially at the same rate at all latitudes in
this layer.  Suppose $\Delta T$ is the pole-equator temperature difference
at the depth $D$.  As we move from this depth $D$ towards the surface,
the overall temperature would keep falling, but $\Delta T$ would still
be the pole-equator temperature difference at the photospheric surface
due to the temperature falling at the same rate at all latitudes.  In
other words, a pole-equator temperature difference at depth $D$ gets
directly mapped to the photospheric surface, even though the overall
temperature keeps falling. If $\Delta T$ is really 2.5 at the photospheric
surface as claimed by Rast, Ortiz, and Meisner (2008), then we have to conclude that
$\Delta T$ at the depth $D$ also should be 2.5 K.

Presumably, the depth $D$ would be in a region where the centrifugal term
is given by (4) and consequently (9) holds.  Taking $\Delta T \approx
2.5$ K, we conclude that the temperature of the isochoric surface at this
depth would be of order
$$T_S \approx 5.0 \times 10^5 \; {\rm K}. \eqno(10)$$
From the standard model of the convection zone (Spruit, 1974; Bahcall
and Ulrich, 1988), we note that the temperature would have such a value at
a radial distance of about $0.91 R_{\odot}$ from the solar centre, which
gives a depth of  
$$D \approx  63,000 \; {\rm km} \eqno(11)$$
below the photospheric surface. Our contention is that, if the pole-equator
temperature difference at the solar surface is really 2.5 K, then
the pole-equator temperature would continue to remain approximately 2.5 K 
till this depth 63,000 km, in spite of the overall temperature
changing by 2 orders of magnitude between the solar surface and this
layer at depth 63,000 km. If $\Delta T$ remains the same but $T$ keeps
falling as we move radially outward in this layer, certainly $\Delta T/T$
would keep increasing, making the thermal wind term given by (8) stronger
and stronger.  To have an estimate of this term, we may take the temperature
of an intermediate layer to calculate the thermal wind term.  Taking 
$T_S \approx 5 \times 10^4$ K, the thermal wind term given by (8) turns
out to be
$$\frac{1}{r} \frac{g}{\gamma \, C_V} \frac{\pa S}{\pa \theta} \approx 1.4 \times 10^{-11}  
\; {\rm s}^{-2}.
\eqno(12)$$

The value of the thermal wind term in this top layer given by (12) is clearly
larger than the centrifugal term in the interior of the convection zone as given
by (4).  If we want the centrifugal term and the thermal wind term to balance
each other even in this thin layer, then clearly $d \Omega^2/ d z$ has to be larger
in this layer, showing the necessity of a near-surface shear layer.  We have already
pointed out that the value of $\Omega/ 2 \pi$ is 420 nHz at the point where a
vertical straight line starting from the bottom of the convection zone in the equatorial
plane intersects the bottom of the near-surface layer we are considering.  If
$\Omega_{\rm top}/ 2 \pi$ is the value of the angular frequency at the point
where this vertical straight line extended further would meet the solar surface,
then the centrifugal term in this near-surface layer is given by
$$r \sin \theta \frac{\pa}{\pa z} \Omega^2 \approx - 
\frac{R_{\odot}}{D} \left[(420)^2 - \left(\frac{\Omega_{\rm top}}{2 \pi}\right)^2 \right] 
\times (2 \pi 10^{-9})^2 \; {\rm s}^{-2}. \eqno(13)$$
If we require that the centrifugal term and thermal wind term should be comparable
even within this near-surface layer, then we have to equate (12) and (13). On using
the value of $D$ given by (11), this gives
$$\frac{\Omega_{\rm top}}{2 \pi} \approx 380 \; {\rm nHz}. \eqno(14)$$
Looking at Figure~1 of Howe (2009), we note that the observational
value is about 400 nHz. In other words, the jump in the value of $\Omega/2 \pi\;$according to our theoretical estimates is $\approx 40$ nHz, whereas the observational
jump is $\approx 20$ nHz.

We thus find that our simple considerations give various important characteristics
of the near-surface layer---such as its depth $D$ and the jump in $\Omega/2 \pi$
within this layer---within a factor about 2 of the observational values.

\section{Conclusion}

It is generally believed that the centrifugal force term and thermal wind term
approximately balance each other within the interior of the convection zone.  From this
balance condition, we estimate the pole-equator temperature difference.  We argue that
this temperature difference is appropriate for a layer at some depth below the surface
and that this temperature difference remains the same as we move though the 
near-surface layer towards the surface, although the overall temperature keeps
falling.  Whether the thermal wind balance should hold within the near-surface layer
as well is not a settled question.  Hotta, Rempel, and Yokoyama (2015) argued that the centrifugal
force term in this layer should be balanced by the turbulent dissipation term.  We
point out that the thermal wind term will become larger in this layer and propose 
the alternative viewpoint that this term will have to be balanced by the centrifugal
force term.  This suggests that the near-surface layer has to be a layer of
strong differential rotation.  The various characteristics of this layer which we
infer agree with the properties of the near-surface shear layer measured by
helioseismology.

We point out that we have not considered magnetic forces in our discussion.  Although
the magnetic forces are believed to drive the torsional oscillations (Chakraborty,
Chatterjee, and Choudhuri, 2009) and the variations of the \MC\ with the 
solar cycle (Hazra and Choudhuri
2017), they presumably are not important in determining the mean characteristics of
the large-scale flows.

The value of the pole-equator temperature difference at the surface is crucial in
estimating the characteristics of the near-surface shear layer.  On using the value
2.5 K, as reported by Rast, Ortiz, and Meisner (2008), we find that the characteristics of this
layer which we obtain are in reasonable agreement with measurements from helioseismology.
In case there is no pole-equator temperature difference at the surface, then our idea
will clearly not work.  Due to the difficulties in treating the near-surface layer
realistically in numerical simulations, it is not possible to draw any firm conclusion
about this temperature difference from simulations (Bidya Karak, private communication).
We hope that the temperature difference at the solar surface will be firmly
established by independent measurements of different groups in the near future.

Finally, we point out that in this paper we do not address the important question
of the possible angular momentum transport mechanisms which must be needed for
sustaining the
near-surface shear layer.  Inclusion of such dynamical considerations will be
essential for developing a fully self-consistent model of the near-surface
shear layer.  In the present paper, our aim is to show that the mere existence of a 
pole-equator temperature difference would suggest the existence of a near-surface
layer within which a strong shear would be needed for balancing the thermal wind
term.

\bigskip

{\bf Acknowledgements.}  I thank Leonid Kitchatinov and Bidya Karak for valuable
discussions.  My research was supported in part by a J.C.\ Bose Fellowship awarded
by the Department of Science and Technology, Government of India.

\section*{References}

Bahcall, J. N., and Ulrich, R. K.: 1988,  Rev. Mod. Phys. 60, 297.

{

\leftskip=20pt
\parindent=-\leftskip
Balbus, S. A.,  Bonart, J., Latter, H.N., Weiss, N. O.: 2009, MNRAS 400, 176.

Belvedere, G., and Paterno, L.: 1976,  Solar Phys.\ 47, 525.

Brown, B. P., Browning, M. K., Brun, A. S., Miesch, M. S., Toomre, J.: 2010,
Astrophys. J.\ 711, 424.

Chakraborty, S., Choudhuri, A. R., Chatterjee, P.: 2009, Phys. Rev. Lett.\ 102, 041102.

Choudhuri, A.R.: 2020, in preparation.

Durney, B. R., and Roxburgh, I. W.: 1971, Solar Phys.\ 16, 3. 

Gastine, T., Yadav, R. K., Morin, J., Reiners, A., Wicht, J.: 2014, MNRAS 438,
L76.

Guerrero, G., Smolarkiewicz, P. K., Kosovichev, A. G., Mansour, N. N.:
2013, Astrophys. J.\ 779, 176.

Hazra, G. and Choudhuri, A. R.: 2017, MNRAS 472, 2728.

Hotta, H., Rempel, M., Yokoyama, T.: 2015, Astrophys. J.\ 798, 51.

Howe, R.: 2009, Living Rev. Solar Phys., 6, 1.

Karak, B. B., Jiang, J. Miesch, M. S., Charbonneau, P., Choudhuri, A. R.: 2014, Space Sci. Rev.\ 186, 561.

Kippenhahn, R., and Weigert, A.: 1990, Stellar Structure and Evolution, Springer-Verlag.

Kitchatinov, L. L.: 2013, in A. G. Kosovichev, E. de Gouveia Dal Pino and Y. Yan (eds.),
Solar and Astrophysical Dynamos and Magnetic Activity: IAU Symposium 294, 399.  

Kitchatinov, L. L., and R\"udiger, G.: 1995, Astron. Astrophys.\ 299, 446.

Kuhn, J. R., Libbrecht, K. G., Dicke, R. H.: 1988, Science 242, 908.

Rast, M.P.,  Ortiz, A., Meisner, R. W.: 2008, Astrophys. J.\  673, 1209.

R\"udiger, G.: 1989, Differential Rotation and Stellar Convection, Gordon \& Breach.

Spruit, H.C.: 1974, Solar Phys.\ 34, 277.

}

\end{document}